# A photo-mechanical coupling theory for photoisomerization hydrogel considering the distribution state of molecular chains

Xinyu Liu, Qingsheng Yang*, Wei Rao*

Dep. of Engineering Mechanics, Beijing University of Technology, Beijing 100124

* Corresponding author: Dr. QS Yang, E-mail address: qsyang@bjut.edu.cn

Dr. Wei Rao, E-mail address: rw@bjut.edu.cn

**Abstract:** Owing to the possibility of controlling its specific mechanical behaviors uptaken by irradiated by light at particular wavelengths, the photoisomerization hydrogels have a broad range of potential applications. A theory connecting the optical excitation to mechanical behavior is essential to precisely control the photo-mechanical behaviors of the hydrogel. In this work, a photo-mechanical coupling theory is developed to describe the photo-mechanical responses of photoisomerization hydrogels within the framework of finite deformation continuum thermodynamics. In the model, the deformation gradient is decomposed into two parts to effectively model the light-induced deformation and the elastic one. To consider the effect of the optical excitation on mechanical behaviors, we first investigate the transporting mechanism of light in hydrogel, as well as the photochemical reaction process; and we then explore the disturbance of light irradiation on the equilibrium of the thermodynamic system of hydrogel, as well as the relationship of conformational entropy of hydrogel network with the photochemical reaction; finally, based on the entropy elasticity theory, we propose a new free energy function of the photosensitive hydrogel to consider the effect of molecular chain distribution evolution on the stiffness of the hydrogel network. With the implementation of the proposed model, we study the photo-mechanical behaviors and mechanical properties of photoisomerization hydrogels. The present research is helpful for understanding the multi-field coupling behaviors of the photosensitive hydrogel, and then providing guidelines for the application of photoisomerization hydrogel.

## 1. Introduction

As a representative of the emerging colloidal intelligent soft materials, hydrogels exhibit good chemo-mechanical coupling characteristics (Ji et al., 2020; Sang et al., 2021; Wu et al., 2022). By controlling the external environmental variables, hydrogels can achieve a series of specific functions (Li et al., 2019). Thus, they have shown excellent application potential and broad application prospects in the fields, such as energy (Kim et al., 2015), environment (Zheng et al., 2023), and anti-tumor (Cai et al., 2019; Hu et al., 2022), drug delivery (Costa et al., 2022; Fan et al., 2019), and get more and more attention in recent years (Li and Su, 2018). Recently, some functional groups, which are sensitive to external stimuli, have been introduced into the polymer network of hydrogel (Koetting et al., 2015). As a result, such new-type hydrogels show a unique mass diffusion-chemo-mechanical coupling deformation behavior (Böger et al., 2017) when they are subjected to the action of the factors, such as optical (Lin et al., 2021), thermal (Liu et al., 2022; Qin et al., 2022), electrical (Ha et al., 2020), magnetic (Liu et al., 2020) and PH (Fan et al., 2019) etc.. Generally, with the emergence of those stimulus-responsive hydrogel, stronger and stronger intelligence renders in the hydrogels to be a more excellent functional material (Ahmed, 2015; Caló and Khutoryanskiy, 2015).

As a new favorite of stimulus-responsive hydrogel, the photosensitive hydrogels are polymeric networks introduced photosensitive group, which absorb and retain large amounts of water (Wu et al., 2018). Since light possesses the unique merits, such as clean, safe, remote controllability and rapidity, the study on photosensitive hydrogels is in full swing (Lee et al., 2015). Consequently, many kinds of new-type photosensitive hydrogels have sprung up in recent years. According to the photochemical reaction mechanism of photosensitive group, those existing photosensitive hydrogels can be divided into five kinds, including: photoionization (Dehghany et al. 2018), photopolymerization (Peng et al., 2011; Wei et al., 2021; Zhu et al., 2020), photolysis (Tibbitt et al., 2013; Zhu and Bettinger, 2013), photothermal (Brighenti and Cosma, 2022; Jaik et al., 2022; Wang et al., 2021) and photoisomerization hydrogels (Kim et al., 2017; Wang et al., 2019). Correspondingly, the photochemical reactions with different mechanisms have different effects on the microstructures and properties of photosensitive hydrogels. Based on the objects to which the photosensitive group functions, such photosensitive hydrogels can be classified as: (1) affecting the solution, among them, the photoionization hydrogel is one of the representatives (Maleki et al., 2021); (2) influencing the polymer networks (Brown et al., 2018; Yu et al., 2019), and all of photopolymerization, photolysis, photoisomerization hydrogels are belonging to such class. It should be noted that since the photosensitive group functioning to solution only has some effects on the concentration of the solution during the photochemical reaction, the swelling deformations are usually the critical concern of such hydrogels. Instead, the photosensitive group functioning to the polymer network always changes the distribution, cross-linking degree of polymer network during the photochemical reaction (Ma et al., 2014). The light stimulation may not only directly induce deformations of the hydrogel, but also affect the stiffness of polymer network of hydrogels (Accardo and Kalow, 2018). Meanwhile, since the photochemical reaction influences the shape and stiffness, the swelling of such hydrogels is also sensitive to the light stimulation. Genially, their mechanical properties are always strongly dependent on the external light field.

Since the polymerization reaction shows good macroscopic reversibility, the photoisomerization hydrogels receive a warm welcome in the biomedical applications (Pianowski et al., 2016) or the development of sensors (Chen et al., 2018; Debroy et al., 2019). Recently, some researchers have conducted some studies on the deformation mechanism of photoisomerization hydrogel, and found that it is essentially a mutual transformation within isomers of polymers (Rosales et al., 2015). For example, the azobenzene functional group transits from the ground state to the single excited state or triple excited state after receiving photons; as a result, the N=N double bond in a functional group would be driven to undergo the rotation, and the molecular structure of azobenzene would change from trans to cis states (Cembran et al., 2004; Peng et al., 2012). Along with the structural change of azobenzene, the spatial distribution and intermolecular polarity of polymer networks in hydrogels also change. Finally, the changes in macroscopic shape and stiffness of polymer networks are the expression form of photo-mechanical coupling behaviors for photoisomerization hydrogels. Overall, the existing research have clearly revealed the photo-mechanical coupling mechanisms of photoisomerization hydrogels. However, since the deformation mechanisms of photoisomerization hydrogels are very complex, developing predictive continuum models that effectively establish a quantitative relationship between photoisomerization reaction and mechanical properties remains a major challenge of prime scientific and technological importance. Consequently, how to realize the fine control is a truly challenging problem.

Given the great significance, modeling the photo-mechanical coupling behaviors of photosensitive hydrogels has attracted a lot of attention in the last five years, although the photosensitive hydrogels are prepared shortly. So far, several new constitutive models have been proposed to describe deformations of photosensitive hydrogels with different photochemical reaction mechanisms, including: the constitutive model of photoionization hydrogel developed by Dehghany et al. (2018), the constitutive model of photo-polymerization/photolysis hydrogel developed by Zhao et al. (2019), the constitutive model of photoisomerization hydrogel proposed by Xuan and Jin (2019). These constitutive models can effectively capture some photo-mechanical coupling behaviors of photosensitive hydrogels, such as different swelling behaviors displayed by photosensitive hydrogels subjected to the stimulation of different external light fields. However, since the correlations of photochemical reaction on the microstructure evolution (i.e., distribution) of polymer network are lack of attention, a new constitutive model is needed to be further developed for the photo-mechanical coupling behaviors of photoisomerization hydrogels.

To make up the defects, we develop a photo-mechanical constitutive theory for the photo-mechanical responses of photoisomerization hydrogels within the framework of finite deformation continuum thermodynamics. In the model, the effect of photoisomerization reaction on deformations of hydrogel is summarized into three parts, i.e., the light-induced deformation, the disturbance on the equilibrium of thermodynamic system, and the effect of the end-to-end distance of polymer chains. In order to effectively consider such three kinds of mechanisms, we first decompose the deformation into two parts to effectively model the light-induced deformation and the elastic one, and then propose a new thermodynamic framework, finally derive the relationship of conformational entropy of hydrogel network with the photochemical reaction based on the Gaussian chain statistics method. Moreover, based on the entropy elasticity theory, a new free energy function of the photosensitive hydrogel is also given to consider the effect of molecular chain distribution state on the mechanical properties of the hydrogel network. It should be noted that since the photoisomerization reaction is closely related to the photo-mechanical coupling behaviors of hydrogel, a transport function of light is proposed to model the evolution of the light field in hydrogel, and a state variable is introduced to describe the photoisomerization reaction process. Finally, we adopt the model to discuss the photo-mechanical coupling behaviors and the effect of photo-mechanical coupling action on the mechanical properties of photoisomerization hydrogels.

## 2. Constitutive model for photoisomerization hydrogel

Since the introduced photosensitive functional groups are varied, there are different kinds of photoisomerization hydrogels. Consequently, their photo-mechanical mechanisms are not the same, although they are affiliated with photoisomerization hydrogel. In this work, we only focus on the deformations of photoisomerization hydrogel containing the azobenzene functional group. For such photoisomerization hydrogel, the effect of its photochemical reaction on mechanical properties can be summarized as: (1) light-induced deformation, as shown in Fig. 1, (2) the photochemical reaction influencing the distribution of molecular chains, (3) the photochemical reaction disturbing the equilibrium state of thermodynamical system of hydrogel.

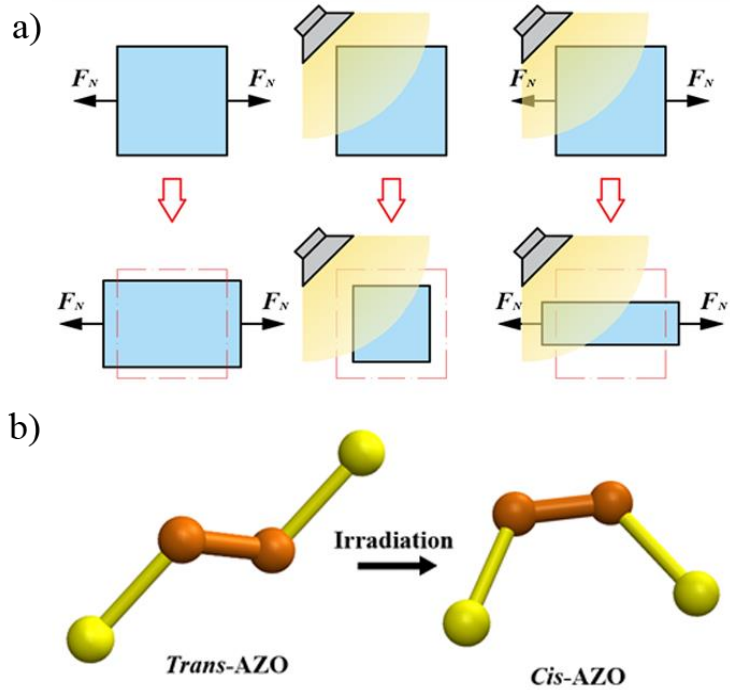

**Figure 1** Deformation mechanism of photoisomerization hydrogel: a) visible macroscopic deformation within irradiated and without irradiated situation; b) change of the AZO segment from trans state to cis state.

In this section, to develop the constitutive model for photoisomerization hydrogel containing the azobenzene functional group, we first propose a thermodynamic framework of finite deformation to consider the coupling effect between light-induced deformation and elastic one and the disturbance of photoisomerization reaction on thermodynamical system, and discuss the function of light transmission and the evolution function of photoisomerization reaction. Then, based on the Gaussian chain statistics method, we derive the quantitative functional relationship between the photoisomerization reaction and molecular chain distribution, and provide a new free energy function for the photoisomerization hydrogel on the basis of the entropy elasticity theory. Finally, a constitutive relationship for photoisomerization hydrogel is proposed by combining the developed thermodynamic framework and free energy function.

## 2.1 Kinetics

Considering a deformable body occupies the region $V_0$, $\boldsymbol{X}$ is the position vector defined in the reference configuration, which means a material point. And $\boldsymbol{x}(\boldsymbol{X},t)$ is the position vector at time $t$ defined in the current configuration, which means a point in the space related to a material point. Then, the deformation gradient $\mathbf{F}$, vector $\boldsymbol{v}$, and velocity gradient $\mathbf{L}$ can be given as

$$\mathbf{F} = \frac{\partial \boldsymbol{x}}{\partial \boldsymbol{X}}, \quad \boldsymbol{v} = \dot{\boldsymbol{x}}, \quad \mathbf{L} = \dot{\mathbf{F}} \cdot \mathbf{F}^{-1} \tag{1}$$

For the photoisomerization hydrogel containing the azobenzene functional group, its deformations are the combining contributions of both elastic deformation and light-induced one when subjecting both the loading and light field. Following the Polar decomposition, the deformation gradient $\mathbf{F}$ can be divided as

$$\mathbf{F} = \mathbf{F}^e \cdot \mathbf{F}^L \tag{2}$$

where $\mathbf{F}^e$ and $\mathbf{F}^L$ are the elastic and light-induced deformation gradients, respectively.

From Eq. (2), it can be obtained that

$$J = \det(\mathbf{F}) = \det(\mathbf{F}^e)\det(\mathbf{F}^L), \quad J^L = \det(\mathbf{F}^e), \quad J^L = \det(\mathbf{F}^L) \tag{3}$$

where $J$, $J^e$ and $J^L$ are the volume Jacobians of total, elastic and ligh induced deformations, respectively.

For most photoisomerization hydrogel, the azobenzene functional groups are always randomly distributed into the polymer network. In other words, the photoisomerization hydrogel doesn't show any polarity. As a result, we can regard that the light-induced deformations can only change the volume of photoisomerization hydrogel. Moreover, each functional group can produce the same light-induced deformation after being excited. That is to say, the degree of light induced deformation is proportional to the number of excited azobenzene functional group, as shown in Fig. 2. Thus, if a part of functional groups are excited, the light-induced deformation gradient can be expressed as

$$\mathbf{F}^L = (f\xi + 1)\mathbf{I} \tag{4}$$

where, $f$ is the ratio of excited functional groups to all the functional groups in hydrogel, and $\xi$ is the light-induced deformation when all the azobenzene functional groups in the hydrogel are excited.

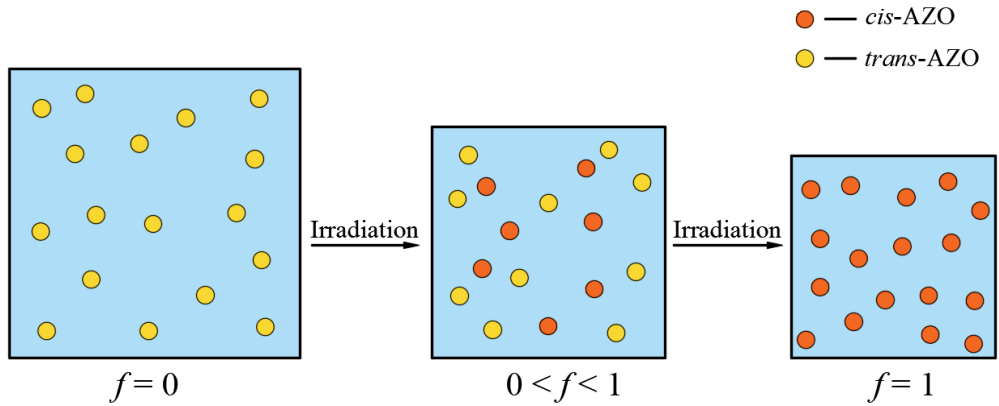

**Figure 2** The trans-AZO changes to cis-AZO within irradiation. The total AZO is a constant. The state variable of $f$ represents the ratio of excited functional groups to all the functional groups in hydrogel. When $f = 0$, the reaction doesn't happen, and when $f = 1$, the reaction is carried out thoroughly.

**2.2 Transport of light**

Since the photoisomerization reaction relies entirely on the local illuminance of light, we should give a light transport function to characterize the evolution of light illuminance when spreading through the hydrogel. To avoid the introduction of the complex electromagnetic field theory, the light field is regarded as the energy field or photon flow of radiation in this work. Moreover, due to the heterogeneity in hydrogel medium, the transport path of light is also very complex. In order to simplify provide a valuable function of light transport, some assumptions are made, including:

(a) the source only provides the monochromatic and unidirectional light;

(b) the deformable body is covered by light completely;

(c) the light always travels in a straight line.

Considering that the hydrogel is irradiated beginning with $t=0$, the microstate of hydrogel and the illuminance of light source are given. The illuminance of light would attenuate with the propagation of light in hydrogel medium, and the attenuation rate is determined by two factors, i.e., (1) the photoisomerization reaction absorbs some

photons; (2) the background absorbs radiations. Then, the light transport function can be given as

$$I(x,t) = I_0 e^{-\alpha(t)l} - I' \quad (5)$$

where $I$ denotes the illuminance of light at time $t$ and location $x$ in body. $I_0$ is the illuminance of light source, $\alpha(t)$ is the light absorbance coefficient, and evaluating with the time because of the change of hydrogel medium during the photoisomerization reaction, $l$ denotes the distance between the point of incidence at the surface and the current location. The first term means the absorption of background. The second term $I'$ is the absorption in photo-reaction.

It should be noted that the strain of hydrogel can exceed 100% before it fails. When it is undergoing a finite deformation, the light irradiating a certain material point may penetrate from different incidence points at the surface of deformable body. Correspondingly, the illuminance of light in a certain material point may change with the increase of deformations. Thus, the distance between the point of incidence at the surface and the current location should be written as

$$l = (x - x_{in}) \cdot n_0 \quad (6)$$

where $n_0$ is a constant vector which is the direction of the light source, $x$ is the spatial location of a certain material at the current configuration, $x_{in}$ is the spatial location of the incidence point at the surface, and changes with the increase of deformation.

Since the source only provides the monochromatic and unidirectional light, and the light always travels in a straight line, it yields

$$\frac{x - x_{in}}{\|x - x_{in}\|} = n_0 \quad (7)$$

Moreover, the incidence point always belongs to the surface of the deformable body. Thus, if the contour function of the deformable body is set as $S(x) = 0$, it can be obtained that

$$S(x_{in}) = 0 \quad (8)$$

Combining Eqs. (7) and (8), $x_{in}$ can be solved. And then from Eq. (6) and (5), the illuminance of light at a certain point can be determined.

**2.3 Photoisomerization reaction**

Irradiated by the light at a specific wavelength (i.e., 365 nm), the azobenzene functional group in hydrogel would be gradually driven from trans to cis states. And then, the distribution of polymer network in hydrogel also changes. In such a reaction, the illuminance of light in the local region is regarded as the only factor determining the reaction rate. Of course, Eq. (9) also shows that the photoisomerization reaction significantly affects the evolution of light illuminance in a local region. As mentioned in section 2.1, the ratio of excited functional groups to all the functional groups in hydrogel is introduced as a state variable of $f$ to describe the process of photoisomerization reaction. Thus, in this section, we will propose the function relationship between $f$ and the illuminance of light in a local region.

To make the argument more precise, suppose that the evolution ratio of excited functional groups to all the functional groups, says $\dot{f}$, is linearly proportional to the illuminance of light in a local region. Furthermore, with the increase of excited functional groups, such an evolution rate would inevitably go down, and once all the azobenzene functional groups are excited, the evolution rate would fall to zero. Accordingly, we assume that

$$\dot{f} = \frac{k_I I}{h\nu}(1-f) \tag{9}$$

where $h$ is the Planck constant, $k_I$ is a constant related to the nature of photoisomerization reaction.

Assuming that there is no excited azobenzene functional group at the initial moment, and then the photoisomerization hydrogels have been continuously irradiated by the light at a specific wavelength after the initial moment, the value of $f$ at time of $t$ can be given as

$$f = 1 - e^{-\frac{k_I \int_0^t I \, d\tau}{h\nu}} \tag{10}$$

Since $I'$ is the light illuminance absorbed by photoisomerization reaction at the unit time, we suppose that $I'$ only depends on photoisomerization reaction rate. Thus, we

can write

$$I' = m\dot{f} \tag{11}$$

where *m* is the materials parameter which is only determined by the nature and content of azobenzene functional group in hydrogel; here, it is a constant.

Substituting Eqs. (10), (11) and (12) into Eq. (5), it can be obtained that

$$I = I_0 e^{-\alpha(t)l}\left(1 + m\frac{k_I}{hv}e^{-\frac{k_I I}{hv}t}\right)^{-1} \tag{12}$$

**2.4 Thermodynamics framework**

The deformation and swelling of photoisomerization hydrogel involve the coupling action of light, chemical, and mechanical fields when subjected the external excitation. Here, a thermodynamically consistent, finite-deformation, photo-chemo-mechanical multi-field coupling, continuum framework is developed for photoisomerization hydrogel. In this framework, we rewrite the mass balance, energy conservation equations and entropy inequality to capture the influence of mass transport in swelling, light propagation in hydrogel, and photoisomerization reaction on the thermodynamic system. Finally, a dissipation inequality is provided.

During the swelling of hydrogels, there is some mass exchange between the hydrogel and external media. Based on the transport theorem, the mass balance equation for hydrogel can be given as

$$\dot{\rho}J + \rho\dot{J} + J\nabla_0 \cdot \boldsymbol{J}_w = 0 \tag{13}$$

where $\rho$ is the density at the current configuration, $\boldsymbol{J}_w$ is the mass flown in at unit area per unit time.

In this process, the volume evolution of hydrogel can be described as

$$J = J^L + v_1 C_1 \tag{14}$$

where $v_1$ denotes the volume of a water molecular, $C_1$ denotes the concentration of water at the reference configuration.

The photo-mechanical coupling behavior of photoisomerization hydrogel is a comprehensive thermodynamic process. In this process, there will be the conversion of chemical energy and thermal one, the conversion of work and energy, the energy and mass exchanging with the environment, etc.. For simplicity, we pay more attention to two mechanisms on the basis of thermodynamic framework. Firstly, the chemical energy is converted to internal energy during the photoreaction. Secondly, during the swelling, the mass transport also changes the internal energy of the system. Thus, based on the energy conservation equation in the classical thermodynamic coupling framework, the new energy conservation equation can be given as

$$\frac{\mathrm{D}}{\mathrm{D}t}\int_{V_0} U \mathrm{d}v = -\int_{\partial V_0} \boldsymbol{J}_q \cdot \boldsymbol{N}_0 \mathrm{d}S + \int_{V_0} \dot{q} \mathrm{d}v + \int_{V_0} \mathbf{S} \colon \dot{\mathbf{F}} \mathrm{d}v - \int_{\partial V_0} E_w \boldsymbol{J}_w \cdot \boldsymbol{N}_0 \mathrm{d}S - \int_{\partial V_0} \boldsymbol{J}_{RB} \cdot \boldsymbol{N}_0 \mathrm{d}S \\ + \int_{V_0} \dot{E}_s \mathrm{d}v \quad (15)$$

where $U$ is the internal energy per unit reference volume, $\mathrm{d}v$ and $\mathrm{d}S$ are the area and volume integral in the reference configuration, respectively; $\boldsymbol{N}_0$ is the unit vector along the face normal, $\mathbf{S} = J\mathbf{T}\mathbf{F}^{-\mathrm{T}}$ is the first Piola–Kirchoff stress, $\mathbf{T}$ is the Cauchy stress. $\boldsymbol{J}_q$ denotes the heat flux, $\boldsymbol{J}_{RB}$ denotes the light flux, $E_w$ denotes the heat supply per unit reference volume due to the photoreaction, $E_s = \hat{E}_s(T, f)$ denotes the heat supply per unit reference volume due to the microstructure change of polymer network during the photoisomerization reaction, $\dot{q}$ is the heat supply rates per unit reference volume.

Based on the Gaussian law, the local energy conservation equation for this system can be given as

$$\dot{U} - \dot{q} - \mathbf{S} \colon \dot{\mathbf{F}} + \nabla_0 \cdot (\boldsymbol{J}_q + E_w \boldsymbol{J}_w + \boldsymbol{J}_{RB}) - \dot{E}_s = 0 \quad (16)$$

The entropy evolution of the system is also determined by many factors in the photo-mechanical coupling behaviors. For example, during the photoisomerization reaction, the distribution of polymer network in hydrogel would have changed. Finally, this would result in a change in the conformational entropy of the system. Moreover, during the swelling, the energy of the system would be exchanged with the mass transport. Thus, on the basis of entropy inequality in the classical thermodynamic coupling framework, the new entropy inequality can be given as

$$\frac{D}{Dt}\int_{v_0} \eta \mathrm{d}v \geq \int_{v_0} \frac{q}{T}\mathrm{d}v - \int_{\partial v_0} \frac{\boldsymbol{J}_q}{T}\cdot \boldsymbol{N}_0 \mathrm{d}S - \int_{\partial v_0} \eta_w \boldsymbol{J}_w \cdot \boldsymbol{N}_0 \mathrm{d}S + \frac{D}{Dt}\int_{v_0} \eta_H \mathrm{d}v \quad (17)$$

where $\eta$ is the entropy per unit reference volume. $T$ is the temperature, $\eta_H$ is the conformational entropy of polymer network per unit reference volume, $\eta_w$ is the entropy of system exchanging with the environment per unit mass.

Based on the Gaussian law, the local entropy inequality for this system can be given as

$$\dot{\eta} - \dot{\eta}_H - \frac{\dot{q}}{T} + \nabla_0 \cdot \left(\frac{\boldsymbol{J}_q}{T} + \eta_w \boldsymbol{J}_w\right) \geq 0 \quad (18)$$

Based on the classical thermodynamic, the free energy per unit reference volume can be respectively defined as

$$W = U - T\eta \quad (19)$$

Combining Eqs. (16), (18) and (19), it yields

$$\boldsymbol{S}:\dot{\boldsymbol{F}} - \dot{T}\eta + \frac{\partial E_s}{\partial f}\dot{f} - \dot{W} - \nabla_0 \cdot \boldsymbol{J}_{RB} - \boldsymbol{J}_q \cdot \nabla_0 T - W_w \nabla_0 \cdot \boldsymbol{J}_w \geq 0 \quad (20)$$

with

$$W_w = E_w - T\eta_w \quad (21)$$

Here, $W_w$ is the free energy of water per unit mass.

## 2.5 Free energy

As we know, the free energy of pure hydrogel can be regarded as a compound made of polymer network and water. According to Flory's theory (1982), the deformations of polymer network and water can be described by the entropic elasticity theory and the assumption of lattice-liked, and the total responses of the hydrogel system are determined by the sum of such two separate processes. For photoisomerization hydrogel, the photoisomerization reaction only directly affects the entropic elasticity process of polymer network. Thus, it only needs to provide a new entropic elastic model for photoisomerization hydrogel to consider the effect of photochemical reaction.

Since the time scale of the photoisomerization reaction is much smaller than that of mechanical responses, the photoisomerization reaction can be regarded as a transient event. For a chain with *m* light-responsive segments, there are three kinds of segment during the photoisomerization reaction, i.e., normal segment with the characteristic length of | *l* |, *cis*-segment with characteristic length of | $l_{cis}$ | and trans-segment with characteristic length of | $l_{trans}$ |, as shown in Fig. 3.

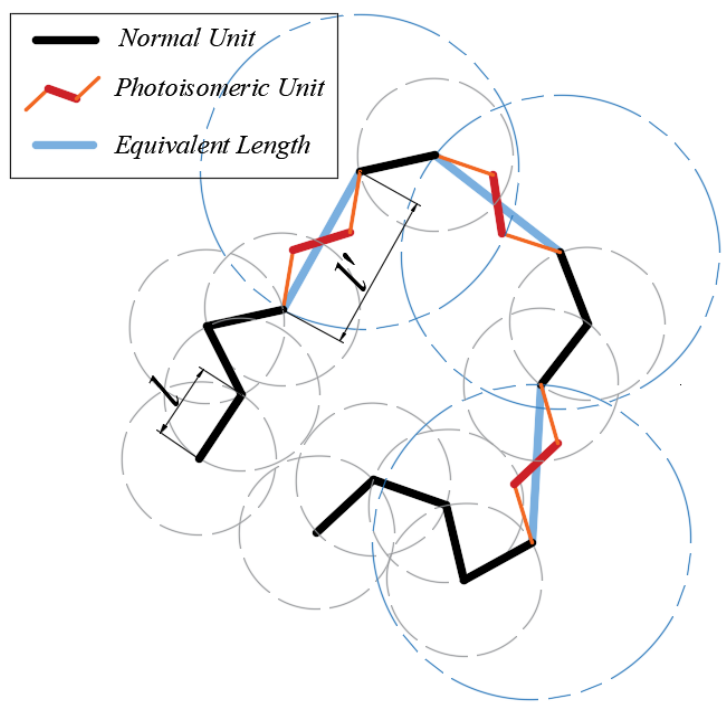

**Figure 3** A single chain contains photoisomerization segments with azobenzene functional group. The segments are all Gaussian chains and can rotate freely on the circles.

For simplicity, the normal segment is considered to be equal in length. Thus, based on the assumption of the Gaussian chain, the probability density of a segment's distribution under the spherical coordinate system is given as

$$\varphi(\boldsymbol{l}) = \frac{1}{4\pi l^2} \delta(|\boldsymbol{l}| - l) \tag{22}$$

where *l* is the length of the segment and *l* is a vector denoting the direction of a segment in space. $\delta(x)$ is the Dirac function.

Based on the Gaussian chain theory, the configuration can be described by the end pitch of chains. Here, since the polymer network has three different segments, the total

end pitch can be regarded as the sum of those three kinds of segments. Thus, we can give

$$h = \sum_{i=1}^{n-m} l_i + \sum_{j=1}^{\alpha} l_{cis\ j} + \sum_{k=1}^{m-\alpha} l_{trans\ k} \qquad (23)$$

where $h$ denotes the total end pitch which is a sum of vectors. $n$ denotes the number of segments, $\alpha$ denotes the number of cis-segments, $m$ is the number of segments embedded in the azobenzene functional group.

Assuming that all the segments embedded in the azobenzene functional group would become the cis-one when all the azobenzene functional groups are excited, the ratio of excited functional groups to all the functional groups in hydrogel can also be defined as

$$f = \frac{\alpha}{m} \qquad (24)$$

Based on the assumption of the Gaussian chain, the rotation of bond angles never suffers the constraint from the neighbor segment. Thus, the probability of each segment distribution is independent. Then, the probability density $\psi$ of configuration is a product of the probability of composing a single chain in the whole segments, i.e.,

$$\psi = \prod_{i=1}^{n-m} \varphi(l_i) \prod_{j=1}^{\alpha} \varphi(l_{cis\ j}) \prod_{k=1}^{m-\alpha} \varphi(l_{trans\ k}) \qquad (25)$$

Since the probability density $\Phi$ of the whole chain can be regarded as the sum of that of all configurations whose end pitch is $h$, the value of probability density $\Phi$ can be given as

$$\Phi(h, n) = \int dl_1 \cdots \int dl_{n-m} \int dl_{cis\ 1} \cdots \int dl_{cis\ \alpha} \int dl_{trans\ 1} \cdots \int dl_{trans\ m-\alpha} \delta\left(\sum_{i=1}^{n-m} l_i + \sum_{j=1}^{\alpha} l_{cis\ j} + \sum_{k=1}^{m-\alpha} l_{trans\ k} - h\right) \psi$$

(26)

To simplify, some constants can be defined as:

1) $\gamma$ denotes the ratio of the number of segments embedded in the azobenzene functional group to all the number of segments in hydrogel, i.e.,

$$\gamma = \frac{m}{n} \qquad (27)$$

2) $k$ denotes the ratio of the length of the cis-segment to the trans-segment ($k < 1$), it

yields

$$k = \frac{|l_{cis}|}{|l_{trans}|} \tag{28}$$

3) $\beta$ denotes the ratio of the length of the trans-segment to the normal segment, it gives

$$\beta = \frac{|l_{trans}|}{|l|} \tag{29}$$

In this work, the hydrogel is assumed to be a homogeneous and isotropic material, so the orientation of the end pitch is in the same probability in any direction. Thus, the influence of orientation on the probability density of the whole chain can be neglected. Then, the distribution function of the length of the end pitch can be given as

$$Y(h,n) = 4\pi h^2 \left[ \frac{3}{2\pi nl^2 B(f)} \right]^{3/2} \exp\left[ -\frac{3h^2}{2nl^2 B(f)} \right] \tag{30}$$

where $B(f)$ is

$$B(f) = \gamma\beta^2 \left[ (k-1)f + 1 \right] + (1-\gamma) \tag{31}$$

According to the assumption of entropic elasticity, the number of microstates is proportional to the probability distribution, and the scale factor is assumed to be $p$. Moreover, since the volume of the hydrogel can change during the deformations, the number of microstates would also change with the change of volume. Thus, the number of microstates $\Omega$ can be given as

$$\Omega = pJ \left[ \frac{3}{2\pi nl^2 B(f)} \right]^{3/2} \exp\left[ -\frac{3h^2}{2nl^2 B(f)} \right] \tag{32}$$

Based on the definition of Boltzmann's entropy, the configurational entropy is

$$\eta_H = \frac{3}{2} k_B \ln\left( \frac{3(pJ)^{2/3}}{2\pi nl^2 B(f)} \right) - \frac{3h^2}{2nl^2 B(f)} \tag{33}$$

where $k_B$ is the Boltzmann constant.

Considering the deformation in the time interval from $t$ to $t+dt$, the end pitch vector changes from $\mathbf{h}$ to $\mathbf{F} \cdot \mathbf{h}$. If the number of chains per reference volume is assumed to be $N$, the evolution of configurational entropy per reference volume can be described as.

$$\Delta \eta_H = \frac{k_B N}{2} \left( 2\ln \frac{J}{J_0} - 3\ln \frac{B(f)}{B(f_0)} - I_1 + 3\frac{B(f)}{B(f_0)} \right) \quad (34)$$

Here, $I_1$ is the first primary invariant of the right Cauchy-Green tensor of total deformation gradient **F**. $J_0$ and $f_0$ are the volume Jacobian and the ratio of excited functional groups at initial time of $t$, respectively.

Assuming that the configuration of hydrogel is in a natural state, the hydrogel undergoes no deformation. Also, there is no activated azobenzene functional group, and all the segments embedding azobenzene functional group are in trans-state. Then based on the entropic elasticity theory, the free energy of polymer network in hydrogel can be given as

$$W_{net} = \frac{k_B NT}{2} \left( I_1 - 2\ln J + 3\ln \frac{B(f)}{B(0)} - 3\frac{B(f)}{B(0)} \right) \quad (35)$$

As mentioned above, the energy of hydrogel includes the contribution of the solution. Here, following the work of Flory and Rehner (1943), the free energy of the solution is given as

$$W_{mix} = k_B T \left[ C_1 \ln \frac{v_1 C_1}{(f\xi+1)^3 + v_1 C_1} + \frac{\chi_h C_1}{(f\xi+1)^3 + v_1 C_1} \right] \quad (36)$$

where $C_1$ is the concentration of water, $\chi_h$ is the Huggins coefficient.

Correspondingly, the total energy of hydrogel can be described by

$$W(\mathbf{F}^e, T, f, C_1) = \frac{k_B NT}{2} \left( I_1 - 2\ln J + 3\ln \frac{B(f)}{B(0)} - 3\frac{B(f)}{B(0)} \right)$$
$$+ k_B T \left[ C_1 \ln \frac{v_1 C_1}{(f\xi+1)^3 + v_1 C_1} + \frac{\chi_h C_1}{(f\xi+1)^3 + v_1 C_1} \right] \quad (37)$$

As we know, the tensile deformation and swelling of hydrogel are always assumed as an isocheimal potential process. Following the function relationship between Gibbs free energy and Helmholtz free energy, another new free energy function is usually defined from Legendre transformation, i.e.,

$$\hat{W}(\mathbf{F}^e, T, f, \mu) = W(\mathbf{F}^e, T, f, C_1) - \mu C_1 \quad (38)$$

where $\mu$ is the chemical potential.

Combining equations (37) and (38), the new free energy function for the hydrogel can be given as

$$\hat{W}\left(\mathbf{F}^{e},T,f,\mu\right)=\frac{k_{B}NT}{2}\left(I_{1}-2\ln J+3\ln\frac{B(f)}{B(0)}-3\frac{B(f)}{B(0)}\right)$$

$$+k_{B}T\left[C_{1}\ln\frac{v_{1}C_{1}}{(f\xi+1)^{3}+v_{1}C_{1}}+\frac{\chi_{h}C_{1}}{(f\xi+1)^{3}+v_{1}C_{1}}\right]-\mu C_{1} \quad (39)$$

Since the total volume of the hydrogel is the sum of the volume of the dry network and the volume of pure liquid solvent, it yields

$$J=(f\xi+1)^{3}+v_{1}C_{1} \quad (40)$$

Substituting Eqs. (40) in (39), it yields

$$\hat{W}\left(\mathbf{F}^{e},T,f,\mu\right)=\frac{k_{B}NT}{2}\left(I_{1}-2\ln J+3\ln\frac{B(f)}{B(0)}-3\frac{B(f)}{B(0)}\right)+\frac{k_{B}T}{v_{1}}\left(J-J^{L}\right)\left(\ln\frac{J^{e}-1}{J^{e}}+\frac{\chi_{h}}{J}-\frac{\mu}{k_{B}T}\right)$$

(41)

### 2.6 Constitutive equation

Combining Eqs.(20) and (39), the local dissipation inequality can be given as

$$\left[(f\xi+1)\mathbf{S}-\frac{\partial\hat{W}}{\partial\mathbf{F}^{e}}\right]:\dot{\mathbf{F}}^{e}+\left(\xi\mathbf{S}:\mathbf{F}^{e}-\frac{\partial\hat{W}}{\partial f}+\frac{\partial E_{s}}{\partial f}\right)\dot{f}-\left(\eta+\frac{\partial\hat{W}}{\partial T}\right)\dot{T}+\frac{\partial\hat{W}}{\partial\mu}\dot{\mu}$$

$$-\nabla_{0}\cdot\mathbf{J}_{RB}-\mathbf{J}_{q}\cdot\nabla_{0}T-F_{w}\nabla_{0}\cdot\mathbf{J}_{w}\geq 0 \quad (42)$$

Since Eq. (42) should always be satisfied with arbitrary values of the elastic strain rate tensor $\dot{\mathbf{F}}^{e}$ and $\dot{T}$, it yields

$$(f\xi+1)\mathbf{S}-\frac{\partial\hat{W}}{\partial\mathbf{F}^{e}}=0 \quad (43)$$

$$\eta+\frac{\partial\hat{W}}{\partial T}=0 \quad (44)$$

Substituting Eqs. (41) into (43), the first Piola-Kirchhoff stress can be given as

$$\mathbf{S}=k_{B}NT\left[(f\xi+1)\mathbf{F}^{e}-\frac{1}{f\xi+1}\mathbf{F}^{e-T}\right]+\frac{k_{B}TJ}{(f\xi+1)v_{1}}\left[\ln\frac{J-J^{L}}{J}+\frac{J^{L}(J+\chi_{h})}{J^{2}}-\frac{\mu}{k_{B}T}\right]\mathbf{F}^{e-T}$$

(45)

Along with Eqs. (43) and (44), the local dissipation inequality can be rewritten in the reference configuration as

$$\left(\xi \mathbf{S}:\mathbf{F}^e - \frac{\partial \hat{W}}{\partial f} + \frac{\partial E_s}{\partial f}\right)\dot{f} + \frac{\partial \hat{W}}{\partial \mu}\dot{\mu} - \nabla_0 \cdot \mathbf{J}_{RB} - \mathbf{J}_q \cdot \nabla_0 T - F_w \nabla_0 \cdot \mathbf{J}_w \geq 0 \tag{46}$$

From Eq. (46), it can be known that the total dissipation can be divided into four parts, i.e., the dissipation resulting from photoisomerization reaction, mass transport, and heat and light flow dissipations. Here, we assume that these four parts of dissipation are independent of each other. Then, the following inequalities can be obtained from Eq. (46), which are stronger than the requirement of non-negative intrinsic dissipation, i.e.,

$$\frac{\partial \hat{W}}{\partial \mu}\dot{\mu} - F_w \nabla_0 \cdot \mathbf{J}_w \geq 0 \tag{47}$$

$$\left(\mathbf{S}:\mathbf{F}^e - \frac{1}{\xi}\frac{\partial \hat{W}}{\partial f} + \frac{1}{\xi}\frac{\partial E_s}{\partial f}\right)\xi \dot{f} \geq 0 \tag{48}$$

$$-\mathbf{J}_q \cdot \nabla_0 T \geq 0 \tag{49}$$

$$-\nabla_0 \cdot \mathbf{J}_{RB} \geq 0 \tag{50}$$

Assuming that the diffusion is an irreversible process, Eq. (47) can be rewritten as

$$\frac{\partial \hat{W}}{\partial \mu}\dot{\mu} - F_w \nabla_0 \cdot \mathbf{J}_w = 0 \tag{51}$$

if Eq. (51) must be satisfied, the chemical potential $\mu$ should be given as

$$\dot{\mu} = \frac{F_w}{J - J^L}\left[\dot{J} - 3\xi(f\xi+1)^2 \dot{f}\right] \tag{52}$$

The detailed derivation has been given in Appendix B.

From Eq.(41), the partial derivative of $W$ hat with $f$ can be

$$\frac{1}{\xi}\frac{\partial \hat{W}}{\partial f} = k_B NT\left[(f\xi+1)I_1^e - \frac{3}{f\xi+1} + \frac{3}{\xi}\frac{\partial B}{\partial f}\left(\frac{1}{B(f)} - \frac{1}{B(0)}\right)\right]$$
$$+ \frac{3k_B T}{v_1}(J^e - 1)(f\xi+1)^2\left[\ln\frac{J^e - 1}{J^e} - \frac{\mu}{k_B T}\right]$$

$$\tag{53}$$

Substituting Eqs. (53) into inequality (48), it yields

$$\frac{\partial E_s}{\partial f} \geq 3k_B NT \frac{\partial B}{\partial f}\left(\frac{1}{B(f)} - \frac{1}{B(0)}\right) - \frac{3k_B T \xi J^L}{(f\xi+1)v_1}\left[\ln\frac{J^e-1}{J^e} + \frac{J+\chi_h}{J} - \frac{\mu}{k_B T}\right] \quad (54)$$

The above equation shows that the value is not arbitrary. The microstructure evolution of the system displays the obvious directivity during the photoisomerization reaction, and drives the system to reach the maximum value of entropy.

Assuming that the heat flow obeys the Fourier's law of heat conduction, inequality (49) is always satisfied, if it is set as

$$\boldsymbol{J}_q = -k_q \cdot \nabla_0 T \quad (55)$$

where $k_q \geq 0$ denotes the coefficients of thermal conduction.

With the assumption about the light source mentioned above, the flux of background absorption should be

$$\boldsymbol{J}_{RB} = I(\boldsymbol{X},t)\boldsymbol{n}_0 \quad (56)$$

Since the illuminance of light would attenuate the propagation of light in hydrogel medium, it must be satisfied with

$$-\nabla_0 \cdot \boldsymbol{J}_{RB} = -\nabla_0 \cdot \left[I(\boldsymbol{X},t)\boldsymbol{n}_0\right] \geq 0 \quad (57)$$

So far, the constitutive model for hydrogel based on the new thermodynamic framework and free energy function has been established. To model the photo-mechanical coupling behaviors, a time-integration procedure for our developed constitutive model has been proposed in this work.

## 3. Simulation and discussion

As an upstart of stimulus-responsive hydrogel, the mechanical behaviors of polymerization hydrogels have been affected by the polymerization reaction. The existing researches show that such influence can be attributed to the change of molecular chain distribution during the polymerization reaction. However, how much impact on the molecular chain distribution is still lacking study; moreover, the relationship between such change of molecular chain distribution and mechanical properties is still unknown. To deeply reveal the photo-mechanical coupling mechanisms of polymerization hydrogel, its tensile deformations and swelling are systematically numerically simulated. Based on the results, the photo-mechanical behaviors and the network evolution rule of photoisomerization hydrogels are discussed.

In section 2, we have given the function relationship between the first Piola-Krichhoff stress and deformation gradient. In fact, during the finite deformation, the true stress and true strain are adopted to characterize the mechanical responses. Thus, here we reformulate the expression of Cauchy stress, i.e.,

$$\boldsymbol{\sigma} = \frac{k_B NT}{J}\left[(f\xi+1)^2 \mathbf{F}^e \mathbf{F}^{eT} - \mathbf{I}\right] + \frac{k_B T}{v_1}\left(\ln\frac{J^e-1}{J^e} + \frac{J+\chi_h}{JJ^e} - \frac{\mu}{k_B T}\right)\mathbf{I} \tag{58}$$

It should be noted that there must be an initial swelling behavior since the hydrogel is a soft colloidal material made up of polymer network swelling in water. That is to say, there is an initial volume strain for hydrogel. Here, we assume that the initial swelling is determined, the initial deformation gradient can be given as

$$\mathbf{F}^0 = \begin{bmatrix} \lambda_0 & 0 & 0 \\ 0 & \lambda_0 & 0 \\ 0 & 0 & \lambda_0 \end{bmatrix} \tag{59}$$

where $\lambda_0$ is the initial elongation, and here it is set as $\lambda_0 = 3.39$.

Correspondingly, the chemical potential for hydrogel with initial swelling can be given as

$$\mu = k_B NT v_1 \frac{\lambda_0^2 - 1}{\lambda_0^3} + k_B T\left(\ln\frac{\lambda_0^3-1}{\lambda_0^3} + \frac{\lambda_0^3 + \chi_h}{\lambda_0^6}\right) \tag{60}$$

Since the chemical potential is a single-value function of swelling ratio, we adopt the swelling ratio to characterize the chemical potential. Moreover, for simplicity, the swelling of hydrogel is always regarded as a quasi-static process since the swelling rate is very slow and the external chemical potential always keeps at a constant value.

### 3.1 Determination of material parameters

In this work, since we only pay attention to mechanical responses at the room temperature, the temperature is set as $T = 293$ K here. Since the photoisomerization hydrogel is prepared by introducing the azobenzene functional group into the pure hydrogel, some parameters independent of the photoisomerization reaction can be determined by referring to those in classical pure hydrogel. Here, some parameters are set to the same value as those in the work of Hong et al. (2009), i.e., Huggins parameter is $\chi_h$ = 0.1, the volume of water molecule is $v_1$ = 1×10$^{-28}$ mm$^3$ and the density of chains is $N$ = 1×10$^{25}$ 1/mm$^3$. In the work of Dehghany (2018), the range of illuminance is set from 10$^{-3}$ W/m$^2$ to 10$^4$ W/m$^2$. Then according to the work of Lee (2018), we suppose that exciting all the azobenzene functional group needs to be irradiated by the light with the illuminance of 1000 W/m$^2$ for one hour; thus, the constant $k_I$ can be given as 1×10$^{-26}$ m$^2$ approximately. Because the photoisomerization reaction occurs only if the wavelength of light is 380 nm, the frequency of light is set as $v$ = 1×10$^{14}$ 1/s in this work. From Takashima's (2012) work, when the photoisomerization reaction reacted utterly, the volume of hydrogel would is reduced to 30%. Thus, it is set as $\xi$ = -0.3. All the values for the material parameters used in this work are listed in table 1 unless noted otherwise.

Table 1 Material parameters for the photoisomerization hydrogel using in this work

| Parameter | Symbol | Value | Reference |
|---|---|---|---|
| Boltzmann constant (J/K) | $k_B$ | 1.38×10$^{-23}$ | |
| Photoreaction parameter (m$^2$) | $k_I$ | 1×10$^{-26}$ | Lee et al. (2018) |
| Planck constant (J·s) | $h$ | 6.62×10$^{-34}$ | |
| Temperature (K) | $T$ | 293 | |
| Huggins parameter | $\chi_h$ | 0.1 | Hong et al. (2009) |
| Frequency of light (1/s) | $v$ | 1×10$^{14}$ | |
| Molecular volume (mm$^3$) | $v_1$ | 1×10$^{-28}$ | Hong et al. (2009) |
| Chain density (1/mm$^3$) | $N$ | 1×10$^{25}$ | Hong et al. (2009) |
| Maximum light-induced deformation | $\xi$ | -0.3 | Takashima et al. (2012) |

## 3.2 Tensile deformations of photoisomerization hydrogel

To investigate the effect of photoisomerization reaction on the deformations, we systematically simulate the deformations of photoisomerization hydrogels subjected the mechanical loading and light irradiation. Based on the loading order, these simulations can be divided into three kinds, including: (1) first adopt the light field to irradiate, then remove the light field, and finally apply the mechanical loading; (2) apply the tensile loading firstly until the elongation reaches a certain value, then hold the constant elongation and adopt the light field to irradiate; (3) apply the mechanical loading and light field simultaneously. In the first simulation, the light illuminances at the first stage are 500 W/m$^2$, 1000 W/m$^2$ and 2000 W/m$^2$, the illumination time is 1800s. In the second simulation, the specific elongation is 8.475, the light illuminances at the second stage are 500 W/m$^2$, 1000 W/m$^2$ and 2000 W/m$^2$, and the illumination time is 6000s. In the third simulation, the tensile strain rate is set as 0.001s$^{-1}$, the light illuminances are also 500 W/m$^2$, 1000 W/m$^2$, 2000 W/m$^2$. Next, we will study the deformation behaviors based on the simulated results.

Fig. 4 shows the predicted results in the first simulation. From the evolutionary trends of tensile stress-strain curves, the hydrogel still exhibits an apparent hyperelastic behavior after it is irradiated by light with different illuminance. In such a process, with the increase of tensile deformation, the polymer network of hydrogel would tighten on the micro-level, and the stiffness of hydrogel increase on the macro-level. On the whole, the photoisomerization reaction at the first stage has no effect on the basic phenomena of tensile deformation. However, if we quantitatively compare the tensile stress-strain curves of hydrogels that have been irradiated by light with different illuminance, there are some significant differences between those curves. It is found that the differences between stiffness (or stress level) of hydrogel exposed to different illuminance of light at the same strain level increase obviously with the increase of tensile deformation. Based on the existing theories, it is not hard to explain this difference. In fact, the photoisomerization reaction at the first stage can make the polymer network be tightened; and the larger the illuminance of light is, the stronger the tightening degree of polymer network in hydrogel. Thus, for the hydrogel exposed to a larger illuminance of light, continuing to tighten the polymer network in hydrogel would become more difficult. In other words, the tightening degree of polymer network in hydrogel exposed to a larger illuminance of light is more extensive with the increase of tensile deformation, and the difference in stiffness would become more significant. It should be noted that the photoisomerization reaction always results in the shrinking of hydrogel, but the tightened polymer network would continue to tighten during the tensile deformation. Thus, the deformation induced by the photoisomerization reaction is not equivalent to that induced by mechanical loading. And the mechanical loading cannot eliminate the effect of photoisomerization reaction on the mechanical properties of hydrogel. Overall, the illumination at the first stage has a significant effect on the following tensile deformation.

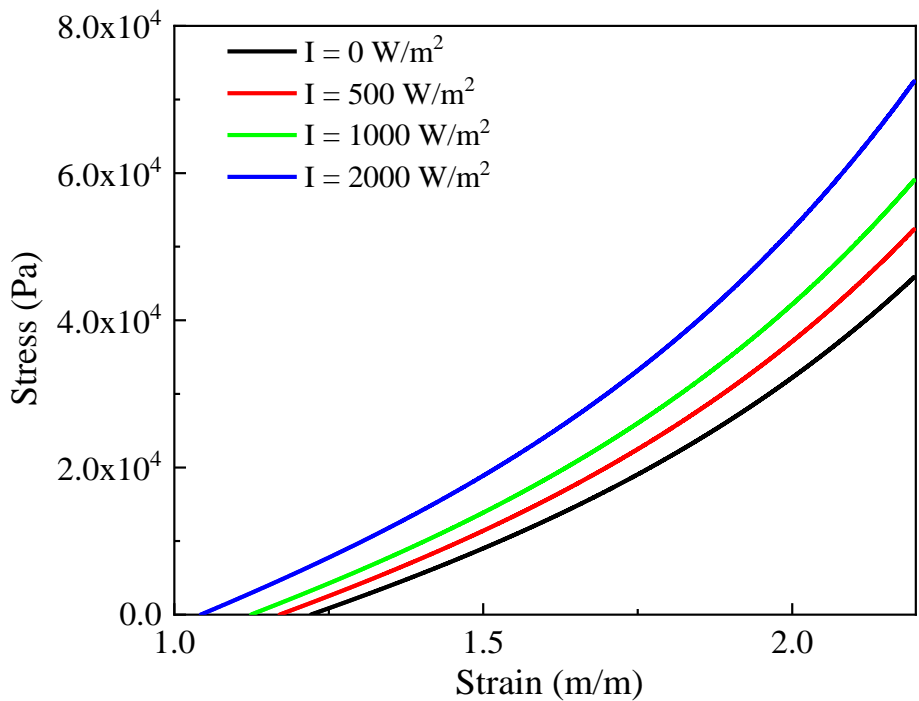

**Figure 4** The relationship between the stress and strain of photoisomerization hydrogels after being irradiated by light with different illuminance for a certain time (1800s)

Fig. 5 shows the predicted results in the second simulation. Notably, in order to investigate the effect of initial tensile deformation on the following light-induced deformation, we also simulate the light-induced deformation of hydrogel without initial tensile deformation. From Fig. 5, it is found that the illuminance of light has a significant effect on the light-induced deformation. The larger the illuminance of light is, the more obvious the increase of stress is with the increase of light-induced deformation. However, with the number of activated azobenzene functional groups increasing, the photoisomerization reaction rate would decrease, and the stress level would tend to be a saturation value. Moreover, if the illumination time is large enough, even the light with the small illuminance can still make the stress reach the saturation value. Since this conclusion is easy to be accepted, we will give a more explanation. More significantly, we find that the initial tensile deformation has some effect on the following light-induced deformation. Once the hydrogel is subjected to some initial deformation, the stress increment is almost an order of magnitude larger than that of hydrogel without initial tensile deformation during the same illumination time. Thus, it indicates that the tightened polymer network at the initial tensile deformation would not loose during the shrinking of hydrogel resulting from the light-induced deformation. That is to say, the photoisomerization reaction cannot eliminate the effect of mechanical loading on the mechanical properties.

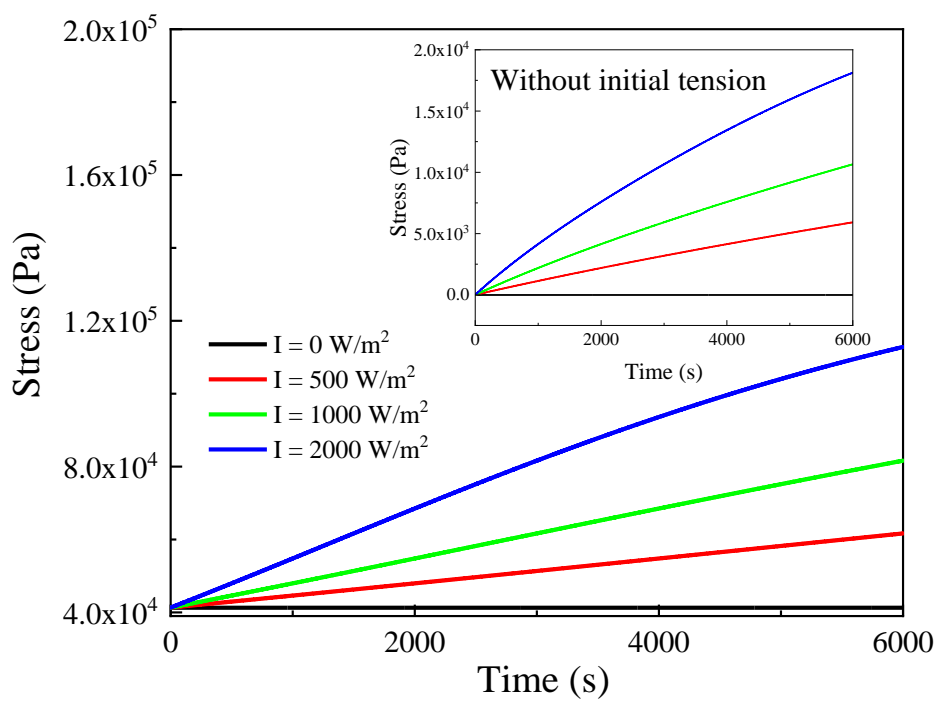

**Figure 5** The stress-applied time curves of photoisomerization hydrogels when being irradiated for 6000 s with different light illuminance within keeping a certain elongation (i.e., $\lambda_1$=8.475).

In order to further discuss the effect of tensile deformation on the light-induced deformation, we also simulated the process of the unloading of the second simulation. Fig. 6 shows the unloading stress-strain curves. From the predicted results, the stress drop rate of hydrogel subjected to a stronger radiation is much more significant during the unloading. Such behavior can be attributed to the hydrogel possessing a larger stiffness when subjected to a stronger radiation. Moreover, it is found that the final elongation of hydrogel subjected to a stronger radiation would be smaller. That is because the stronger radiation can make hydrogel shrink more significantly.

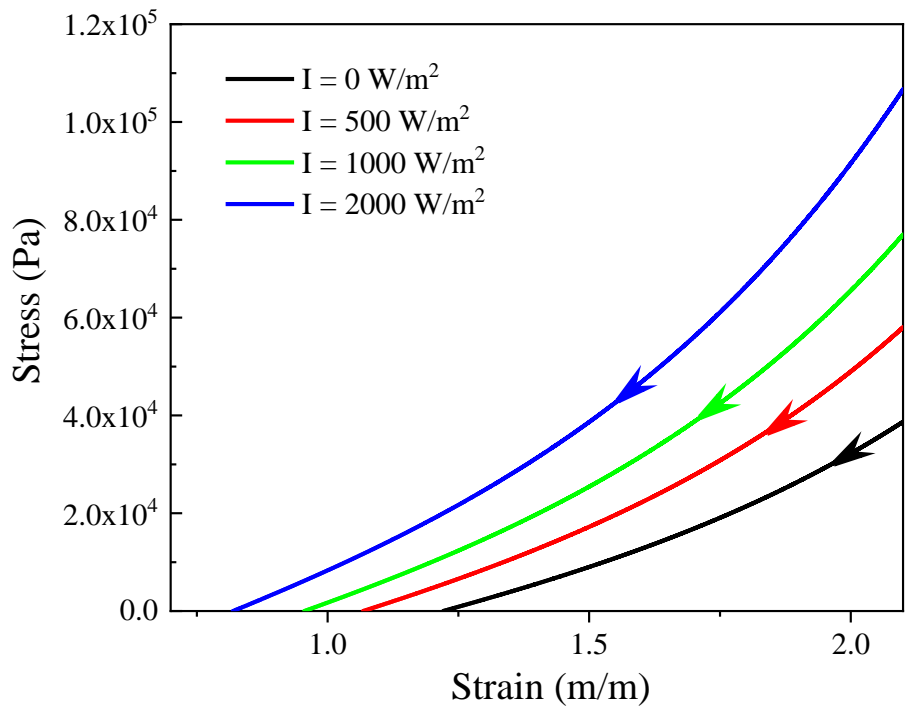

**Figure 6** The relationship between strain and stress of photoisomerization hydrogels when subjecting unload after being irradiated with different light illuminance for 6000 s

Fig. 7 shows the result of the third simulation. Moreover, the unloading stress-strain curves are also provided in Fig. 7. It should be noted that the light field has been removed during the unloading. From Fig. 7, it is found that the hydrogel still displays an apparent superelastic behavior when subjecting the combined action of mechanical loading and light field. However, since the hydrogel has a larger rate of photoisomerization reaction when irradiated by the stronger radiation, the increasing rate of stress level is much faster in the hydrogel irradiated by light with larger illuminance. Moreover, with the increase of loading time, the difference of stress increasing rate becomes larger and larger for hydrogel irradiated by the light with different illuminance. Perhaps more interestingly, it is found that there is a significant difference between the loading curves and unloading ones. By comparison, the stiffness of hydrogel during the unloading is smaller than that of hydrogel at the same strain level during the loading. Moreover, when the hydrogel is subjected to a stronger radiation, the difference in stiffness between the loading process and unloading one tend to be more significant. Of course, these differences can also be ascribed to the effect of photoisomerization reaction. But such behaviors seem to further demonstrate that there is a robust nonlinear coupling effect between the tensile deformation and light-induced deformation. And this coupling effect would continue to affect the following mechanical responses. Thus, the deformation path of hydrogel is always dependent on the loading history.

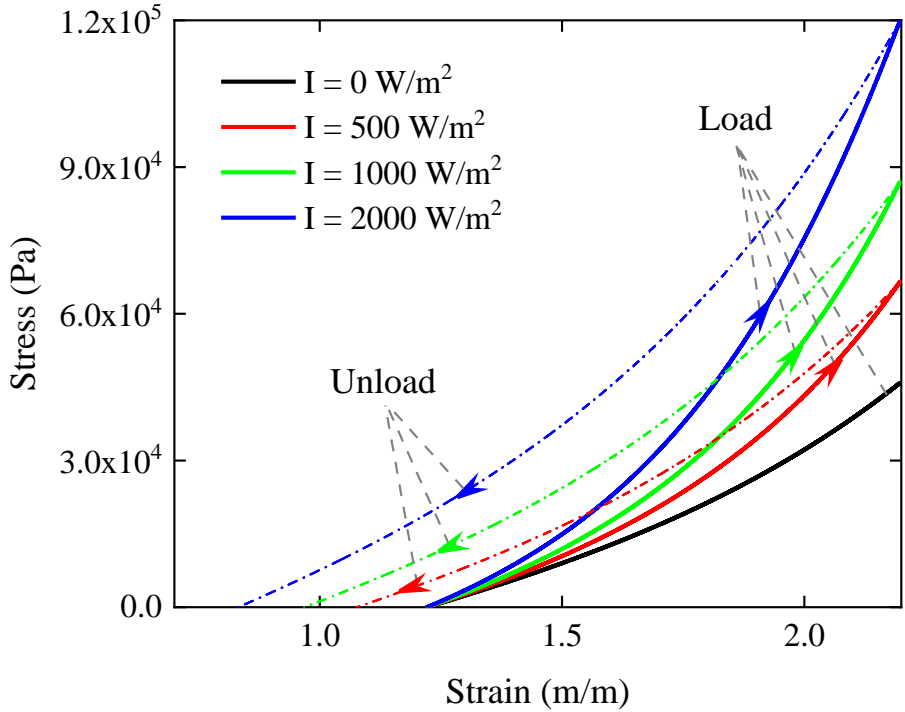

**Figure 7** The relationship between stress and strain during tension and unloading at the elongation rate of 0.001 1/s after the photoisomerization hydrogels are irradiated by different light illuminance

To further discuss the photo-mechanical coupling behaviors of photoisomerization hydrogel, we also model the deformation of hydrogels subjecting the same radiation and the mechanical loading with different rates. Fig. 8 shows the stress-strain responses. From the results, we find that the curves have the same trend as the curves in Fig. 7. Furthermore, the influence of the rate of mechanical loading is the same as that of the illuminance. In fact, with the rate of tensile deformation increasing, the time required for applying the same mechanical loading becomes small, thus the degree of photoisomerization reaction would decrease during deformation. Such an effect is equivalent to reducing the illuminance of light. As a whole, the illuminance and the loading rate have the same effect on the deformation of hydrogel subjected to both the mechanical loading and radiation simultaneously.

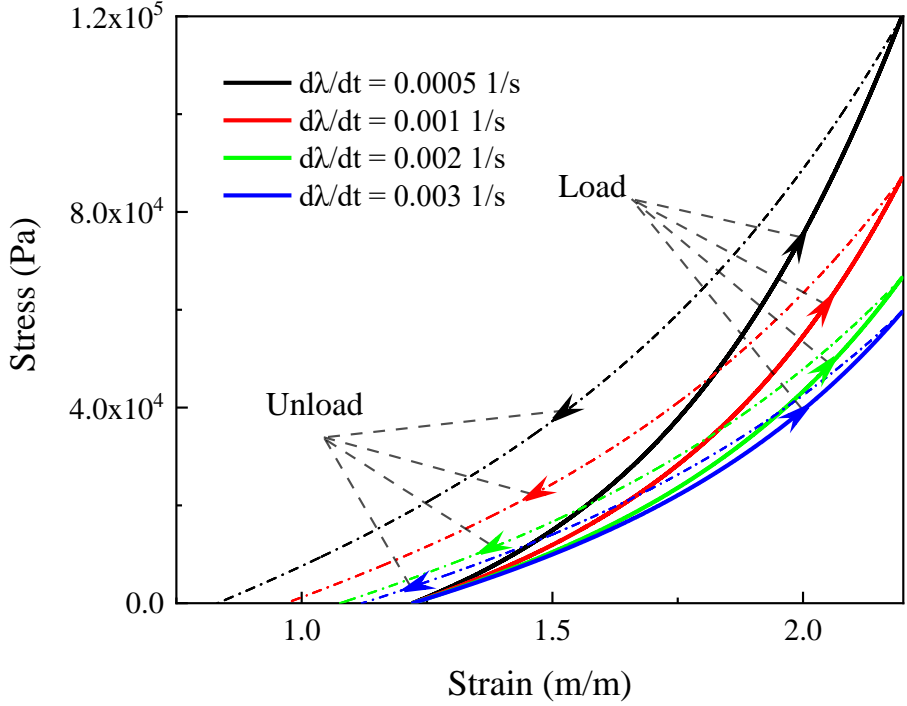

**Figure 8** The relationship between the stress and the strain of photoisomerization hydrogels when it simultaneously subjects the mechanical with different elongation rates and irradiated by the light with the illuminance of 1000 W/m$^2$

### 3.3 Swelling of photoisomerization hydrogel

Free swelling is an essential nature of hydrogel which is different from normal polymers. Here, we investigate the effect of photochemical reactions on the swelling of photoisomerization hydrogel. In simulations, we assume that the initial swelling is still set as $\lambda_0 = 3.39$. It should be noticed that $F_w$ is dependent on the chemical patients $\mu_f$ of external solution. Thus, Eq. (52) can be rewritten as

$$\dot{\mu} = \frac{K_m(\mu - \mu_f)}{J - J^L}\left[\dot{J} - 3\xi(f\xi + 1)^2 \dot{f}\right] \tag{61}$$

Here $K_m$ is a constant associating with the nature of the solution. Through estimating the required time of swelling of pure hydrogel, $K_m$ is approximately set as $10^{-3}$.

To further uncover the photo-mechanical coupling properties, the swelling of photoisomerization hydrogel irradiated by light with different illuminance is simulated.

Fig. 9 shows temporal evolution curves of the total elongation during the swelling. From Fig. 9, it is found that the photoisomerization reaction has a significant effect on the swelling of hydrogel. When the light illuminance is small, the photoisomerization reaction rate is low. In this case, the photoisomerization reaction has a little effect on the initial swelling. Thus, as shown in Fig. 9, when the light illuminance is 50 or 100 W/m$^2$, the swelling of photoisomerization hydrogel is similar to that of pure hydrogel. However, it should be noted that the light-induced deformation is gradually beginning to determine the deformations of hydrogel with the completion of swelling. And the total elongation of hydrogel decreases with the increase of time, and finally reaches a constant value that is smaller than the initial one. Instead, when the light illuminance is relatively large, the responses of hydrogel are different from those of hydrogel irradiated by the light with a small illuminance. In this case, the photoisomerization reaction rate is high, the light-induced deformation is much larger than the swelling. Thus, the macroscopic responses are almost entirely determined by the light-induced deformation. Even if the illuminance is large enough (i.e., 500 and 100 W/m$^2$), the hydrogel cannot display any swelling phenomenon. The total elongation of hydrogel always decreases with the increase of time at the whole process until reaching a saturation value. It should be noted that such saturation value is the same as the constant value displayed by the hydrogel irradiated by light with small illuminance. With the continued increase of illuminance (i.e., 2000 W/m$^2$), the required time of photoisomerization reaction is shorter than that of swelling for the photoisomerization hydrogel. Thus, the light-induced deformation would first end, and then the swelling would dominate the following macroscopic response, and the total elongation of hydrogel would increase with the increase of time in this process. But the final elongation also reaches the same value shown by the hydrogel irradiated by light with small illuminance. On the whole, the swelling process of photoisomerization hydrogel is sensitive to the illuminance of the light field, but the final elongation is always independent of the illuminance of the light field.

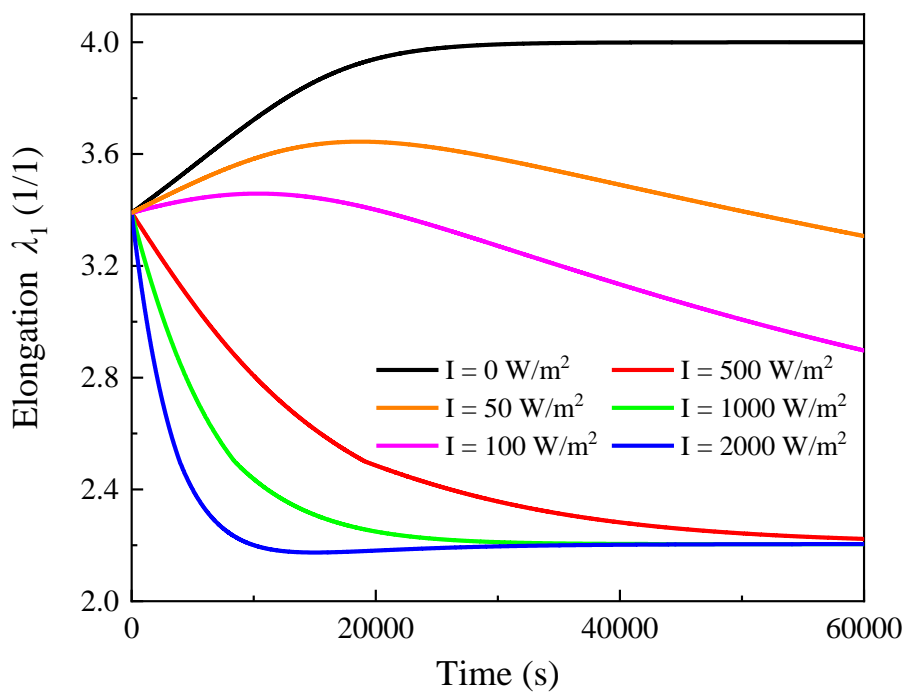

**Figure 9** The relationship between total elongation and time when the photoisomerization hydrogel simultaneously undergoes the swelling at a certain solution and is irradiated by light with different illuminance for some times

**3.4 discussion**

As shown in Figs. 4-9, the illuminance of the light field has a significant effect on the mechanical responses of photoisomerization hydrogel. However, the light field cannot directly influence the deformation behaviors of a pure hydrogel. In fact, that the azobenzene functional group irradiated by light occurs photochemical reaction is the only bridge realizing the photo-mechanical coupling actions. To discuss the photo-mechanical coupling mechanisms, the evaluation rate of the photochemical reaction rate for photoisomerization hydrogel irradiated by light fields with different illuminance has been shown in Fig. 10. From Fig. 10, the rate of photochemical reaction is close to uniform at the certain light illuminance when the illumination time is relatively short; of course, with the increase of illumination time, more and more azobenzene functional groups have been activated, and then the rate of photochemical reaction would gradually slow down until it reaches zero. Of course, as shown in Fig. 10, the rate of photochemical reaction is strongly dependent on the illuminance of light field. With the illuminance of the light field increasing, the rate of photochemical reaction becomes larger at the initial stage, and the number of activated azobenzene functional groups more quickly reaches the point of inflection. Comparing Figs. 5 and 9 with 10, it is easy to demonstrate the above argument which is used to explain the differences in mechanical behaviors for hydrogel irradiated by light with different illuminance in Figs. 5 and 9.

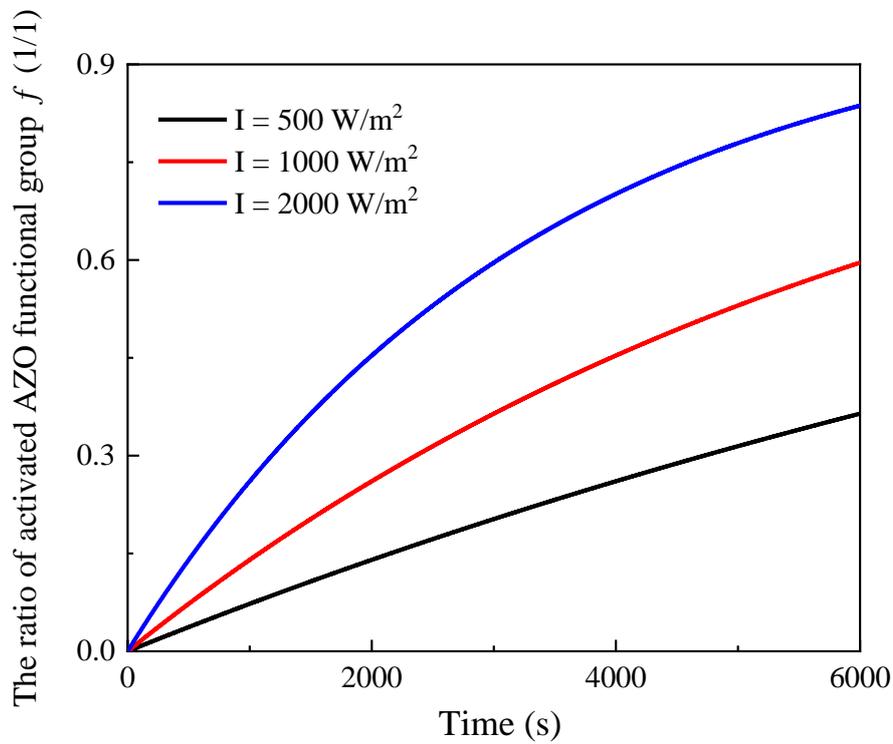

**Figure 10** The evolution of the ratio of excited functional groups to all the functional groups during photoisomerization reaction

As shown in Fig. 10, the degree of photochemical reaction, which affects the mechanical properties (i.e., stiffness) of hydrogel, is significantly dependent on the light illuminance and illumination time. Here, to discuss the photo-mechanical coupling behaviors of photoisomerization hydrogel deeply, Fig. 11 shows the final stiffness of hydrogel when it is firstly irradiated by light with different illumination (i.e., 0, 500,

1000, and 2000W/s) for a certain time (i.e., 1800s, 2400s and 3600s) and then undergoes the tensile deformation until the elongation reaches 8.475 (the black line is the stiffness at the point with the maximum stress each curve in Fig.4). From Fig. 11, it is found that the relationship between of light illuminance and stiffness of hydrogel is approximately linear when the illumination time is relatively short; and with the increase of illumination time, the slope of the linear relationship between stiffness of hydrogel and light illuminance is larger. Comparing Fig. 11 with Fig. 10, we can conclude that all of such appearances are determined by the rate of photochemical reaction, and the argument adopted to explain the differences of curves in Fig. 4 can be further illustrated.

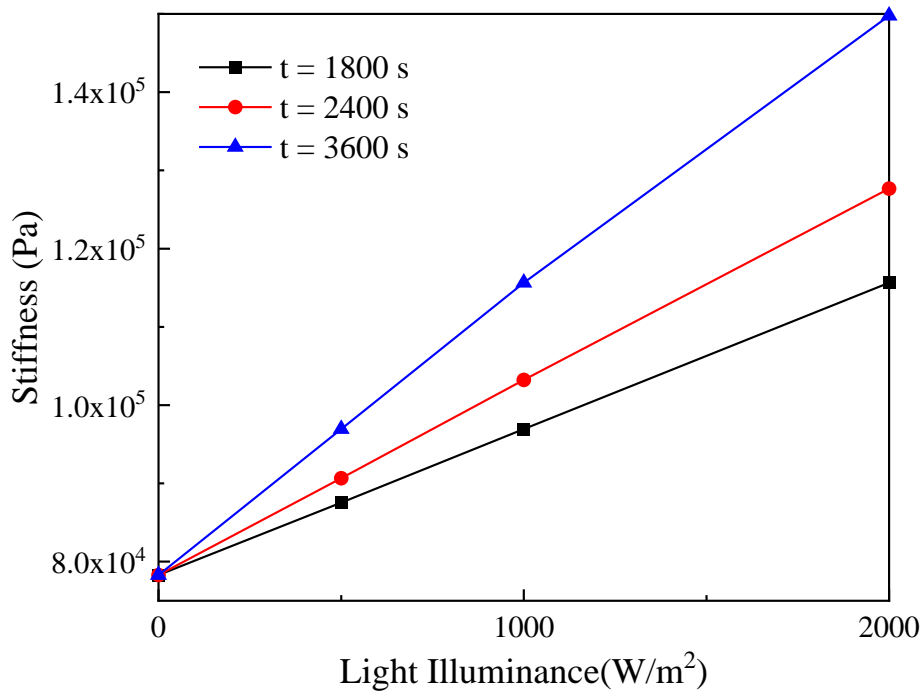

**Figure 11** Stiffness of hydrogels photoisomerization hydrogels when it is firstly irradiated by light with different illuminance for different time (i.e., 1800s, 2400s and 3600s) and then stretched to elongation of a specific value (i.e., $\lambda_1$=8.475).

Furthermore, Fig. 12 shows the stiffness of hydrogel when it is firstly stretched to elongation of some specific values (i.e., 1.5, 2 and 2.5) and then irradiated by the light with different illuminance for 6000 s. From Fig. 12, it finds that the final stiffness of hydrogel still has an approximately linear relationship with the illuminance even if we firstly apply the mechanical loading and then adopt the light field to irradiate. The basic trend of curves in Fig. 12 is the same as that in Fig. 11. However, it should be noted that the differences between the curves in Fig. 12 are more apparent than those in Fig.11. This may indicate that the effect of the initial tensile deformation on the following light-induced deformation is smaller than the effect of initial light-induced deformation on following tensile deformation. Thus, from the results, we conclude that the mechanical responses of photoisomerization hydrogel are dependent on the loading history.

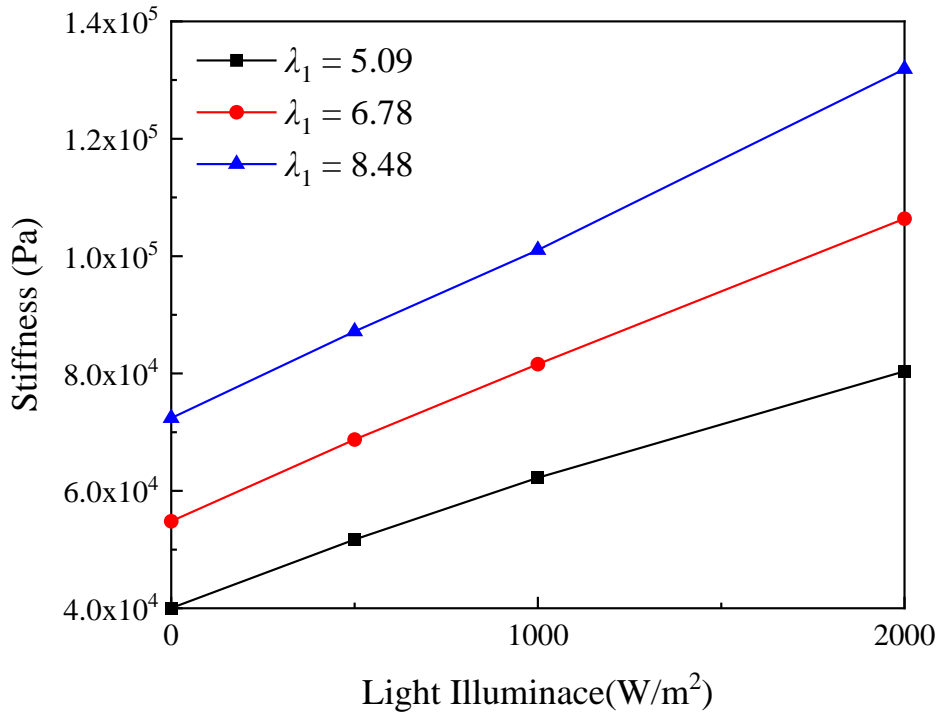

**Figure 12** Stiffness of photoisomerization hydrogel when it is firstly stretched to elongation of some certain values (i.e., 5.09, 6.78 and 8.48) and then irradiated by the light with different illuminance for 6000 s rent light illuminance when irradiated for 6000 s keeping different elongation.

## 4. Conclusion

In this work, we develop a thermodynamic framework of finite deformation that can reasonably capture the coupling effect between light-induced deformation and elastic one and the disturbance of photoisomerization reaction on a thermodynamical system. Moreover, a new free energy function, which can effectively characterize the effect of photoisomerization reaction on the distribution of polymer networks, is proposed. And based on the new thermodynamic framework and free energy function, a photo-mechanical coupling theory is developed, and the photo-mechanical coupling behaviors can be well described by the developed model.

Through simulations, our main findings on the deformation behaviors and mechanism of photoisomerization hydrogel include:

(1) Since there is a strong nonlinear photo-mechanical coupling effect, the tensile deformation and light-induced one of the photoisomerization hydrogels are dependent on the loading history;

(2) The light illuminance has a significant influence on the process of swelling of photoisomerization hydrogel, while doesn't affect on the final swelling degree;

(3) The mechanical load and the light field have different effects on the stiffness of photoisomerization hydrogel. Thus, the effect of mechanical loading on mechanical properties is not equivalent to the effect of photoisomerization reaction.

## Acknowledgement

This work is financially supported by the NSFC (No. 12002342)

## Appendix A. Simplify the inequality of volume fraction dot

The first primary invariant of the right Cauchy-Green tensor of total deformation gradient is defined as

$$I_1^e = \mathbf{F}^e : \mathbf{F}^e \tag{A.1}$$

$$\mathbf{F}^{e-\mathrm{T}} : \mathbf{F}^e = 3 \tag{A.2}$$

Substituting the Eqs. (A.1) and (A.2) into (45), the term $\mathbf{S}:\mathbf{F}^e$ in Eq.(48) can be given as

$$\mathbf{S}:\mathbf{F}^e = k_B NT\left[(f\xi+1)I_1^e - \frac{3}{f\xi+1}\right] + \frac{3k_B TJ}{(f\xi+1)v_1}\left[\ln\frac{J^e-1}{J^e} + \frac{J^L(J+\chi_h)}{J^2}\right] - \frac{3\mu J}{(f\xi+1)v_1}$$

(A.3)

Combining Eqs. (A.3) with (53), we can get

$$\mathbf{S}:\mathbf{F}^e - \frac{1}{\xi}\frac{\partial \hat{W}}{\partial f} = -\frac{3k_B NT}{\xi}\frac{\partial B}{\partial f}\left(\frac{1}{B(f)} - \frac{1}{B(0)}\right) + \frac{3k_B TJ^L}{(f\xi+1)v_1}\left[\ln\frac{J^e-1}{J^e} + \frac{J+\chi_h}{J} - \frac{\mu}{k_B T}\right]$$

(A.4)

And then substituting Eqs. (A.4) into (48), it yields

$$\left\{\frac{1}{\xi}\left[\frac{\partial E_s}{\partial f} - 3k_B NT\frac{\partial B}{\partial f}\left(\frac{1}{B(f)} - \frac{1}{B(0)}\right)\right] + \frac{3k_B TJ^L}{(f\xi+1)v_1}\left[\ln\frac{J^e-1}{J^e} + \frac{J+\chi_h}{J} - \frac{\mu}{k_B T}\right]\right\}\xi\dot{f} \geq 0$$

(A.5)

$\xi$ is a negative value and only consider the one direction reaction which means $\dot{f} > 0$ always is right, then Eq. (54) can be given as

$$\frac{\partial E_s}{\partial f} \geq 3k_B NT\frac{\partial B}{\partial f}\left(\frac{1}{B(f)} - \frac{1}{B(0)}\right) - \frac{3k_B T\xi J^L}{(f\xi+1)v_1}\left[\ln\frac{J^e-1}{J^e} + \frac{J+\chi_h}{J} - \frac{\mu}{k_B T}\right] \quad (54)$$

# Appendix B. Simplify the equation between concentration $C_1$ and chemical potential $\mu$

According to Legendre transformation, Eq. (B.1) can be given as

$$\frac{\partial \hat{W}}{\partial \mu} = -C_1 \quad \text{(B.1)}$$

According to the Fick's law, it yields

$$\dot{C}_1 = -\nabla \cdot \mathbf{J}_w \quad \text{(B.2)}$$

Substituting Eqs. (B.1) and (B.2) into (51), it yields

$$-C_1\dot{\mu} + F_w\dot{C}_1 = 0 \quad \text{(B.3)}$$

From Eq. (40), the rate of $J$ can be

$$\dot{J} = 3\xi(f\xi+1)^2\dot{f} + v_1\dot{C}_1 \quad \text{(B.4)}$$

From Eq.(B.4), the time derivative of $C_1$ can be

$$\dot{C}_1 = \frac{1}{v_1}\dot{J} - \frac{3\xi}{v_1}(f\xi+1)^2 \dot{f} \tag{B.5}$$

Substituting Eqs. (B.5) in (B.3), the derivative time of $\mu$ can be given as

$$\dot{\mu} = \frac{F_w}{v_1 C_1}\dot{J} - \frac{3\xi F_w}{v_1 C_1}(f\xi+1)^2 \dot{f} \tag{B.6}$$

Combining Eqs. (40) and (B.6), the Eq. (52) can be given as

$$\dot{\mu} = \frac{F_w}{J-J^L}\left[\dot{J} - 3\xi(f\xi+1)^2 \dot{f}\right] \tag{B.7}$$